\documentstyle[12pt]{article}

\def\sqr#1{\mathop{\mkern0.5\thinmuskip
\vbox{\hrule\hbox{\vrule
\hskip#1\vrule height#1 width 0pt\vrule}\hrule}
\mkern0.5\thinmuskip}}
\def\Square{\mathchoice{\sqr{6pt}}{\sqr{6pt}}
{\sqr{4pt}}{\sqr{3pt}}}
\def\sq{\Square}

\def\CC{{C\!\!\!\!I}}

\def\tr{{\rm tr\,}}
\def\Tr{{\rm Tr\,}}

\def\det{{\rm det\,}}

\def\log{{\rm log\,}}

\def\vol{{\rm vol\,}}

\def\End{{\rm End}}

\def\be{\begin{equation}}
\def\ee{\end{equation}}
\def\bea{\begin{eqnarray}}
\def\eea{\end{eqnarray}}
\newtheorem{theorem}{Theorem}
\newtheorem{lemma}{Lemma}
\newtheorem{corollary}{Corollary}

\begin{document}

\begin{titlepage}

\null
\vskip-3truecm
\hspace*{7truecm}{\hrulefill}
\par
\vskip-4truemm
\par
\hspace*{7truecm}{\hrulefill}
\par
\vskip5mm
\par
\hspace*{7truecm}{{\large\sf University of Greifswald (July, 1997)}}
\vskip4mm
\par
\hspace*{7truecm}{\hrulefill}
\par
\vskip-4truemm
\par
\hspace*{7truecm}{\hrulefill}
\par
\bigskip
\hspace*{7truecm}hep-th/9707040
\bigskip
\par
\hspace*{7truecm}{\sf to appear in:}\par
\hspace*{7truecm}{\sf }\par
\hspace*{7truecm}{\large\sf J. Mathematical Physics (1998)}\par
\hspace*{7truecm}{\sf }\par
\smallskip

\vskip2truecm

\centerline{\Huge\bf Green functions of }
\medskip
\centerline{\Huge\bf higher-order differential operators}
\bigskip

\bigskip
\bigskip
\centerline{\Large\sf Ivan G. Avramidi
\footnote{On leave of absence from Research Institute for Physics, 
Rostov State University,  Stachki 194, 344104 Rostov-on-Don, Russia}
\footnote{Electronic-mail: iavramid@math.uiowa.edu}}
\bigskip

\centerline{\it Department of Mathematics, University of Greifswald}
\centerline{\it F.-L.-Jahnstr. 15a, D--17489 Greifswald, Germany}
\medskip
\centerline{and}
\medskip
\centerline{\it Department of Mathematics, The University of Iowa}
\centerline{\it 14 MacLean Hall, Iowa City, IA 52242-1419, USA}
\bigskip
\centerline{\today}
\medskip

\vfill

{\narrower
\par
The Green functions of the partial differential operators 
of even order acting on smooth sections of a vector bundle
over a Riemannian manifold 
are investigated via the heat kernel methods.
We study the resolvent of a special class of higher-order 
operators formed by the products of second-order operators 
of Laplace type defined with the help of a  unique
Riemannian metric but with different bundle connections
and potential terms.
The asymptotic expansion of the Green functions near
the diagonal is studied in detail
in any dimension. 
As a by-product a simple criterion for the validity
of the Huygens principle is obtained.
It is shown that all the singularities as well as the 
non-analytic regular parts of the Green functions of
such high-order operators
are expressed in terms of the usual heat kernel
coefficients $a_k$ for a special 
Laplace type second-order operator.

\par}
\bigskip
PACS: 02.30.Jr; 02.40.Vh; 02.90.+p; 04.62.+v 
\end{titlepage}


\section{INTRODUCTION}

The Green functions of partial differential operators
are of great importance in mathematical physics and quantum theory
\cite{hadamard23,dewitt65,gilkey84,barvinsky85}.
In particular, the singularities of the Green functions on the diagonal 
play a crucial role in the renormalization procedure of quantum field theory
\cite{bogolyubov76}.

In this paper we study a special class of differential operators of higher
orders that appear,
in particular, in higher-derivative field theories  (for example, in
higher-derivative quantum gravity
\cite{avr86b}).
Namely, we consider differential operators of higher {\it even} orders, $2N$,
of the form
\be
H=(F_N+m^2)\cdots (F_2+m^2)(F_1+m^2),
\label{1a}
\ee
where $F_i$, ($i=1,\dots,N$) are Laplace type operators, i.e. second-order 
operators with the scalar leading symbol
determined by the metric of the manifold
\be
\sigma_L(F_i)=|\xi|^2,
\ee
where $|\xi|^2\equiv g^{\mu\nu}(x)\xi_\mu\xi_\nu$,
and $m$ is some constant.
In the next section we give precise definition of Laplace type operators 
and describe their properties.
Obviously, the operator $H$ also has a scalar leading symbol
\be
\sigma_L(H)=|\xi|^{2N},
\ee
and is a particular case of the so called minimal higher-derivative operators.

We would like to stress from the beginning that the operators $F_i$, even if of
the same
Laplace type, are {\it different} and, in general, {\it do not commute}. They
only have 
 the same leading, second-derivatives, part, but differ in the first and zero
order parts.

The Green function, or the resolvent kernel, $G_{H-\lambda}(x,x')$
of the operator $H$
is the kernel of the Green, or resolvent, operator $G_{H-\lambda}
=(H-\lambda)^{-1}$ and
is defined by requring it to satisfy the differential equation
\be
(H-\lambda)G_{H-\lambda}(x,x')=\delta(x,x'),
\label{1}
\ee
with $H$ acting on the first argument of the Green function and 
$\delta(x,x')$ being the covariant delta-distribution.

In the present paper we will consider mainly the case of
compact complete manifolds without boundary and assume the metric of the
manifold to be positive definite. Then the Laplace type operators $F_i$
as well as the operator $H$ are {\it elliptic} and eq. (\ref{1}) 
defines a unique Green function.
We will also assume for simplicity the constant $m$ to be
sufficiently large so that all operators $(F_i+m^2)$ and also $H$ 
are {\it  positive}.

However, we would like to make some remarks concerning this subject:
\begin{itemize}
\item[i)]
First, for manifolds with boundary (in elliptic case)
one has to impose additionally some suitable boundary conditions
to make the solution of the eq. (\ref{1}) unique.
However, the singularities of the Green function near the 
diagonal and the asymptotic expansions of it
as $m\to\infty$ do not depend on the boundary conditions.

\item[ii)]
Second, when the metric of the manifold has Minkowski type signature,
${\rm sign}\, g=(-,+,\dots,+)$, the Laplace type operators
$F_i$ are {\it hyperbolic}.
In this case the Green function can be fixed by the Wick rotation,
i.e. the analytic continuation  from the Euclidean case, which
is equivalent to adding an infinitesimal negative imaginary part to the
constant $m^2$, $m^2\to m^2-i\varepsilon$.
In other words, we consider the {\it Feynman propagators}. 
The formulas for the hyperbolic case can be obtained just by analytic
continuation.

\end{itemize}

\section{LAPLACE TYPE OPERATORS}

Let $(M,g)$ be a smooth Riemannian manifold of dimension $d$ with a metric
$g$.
To simplify the exposition we assume the manifold $M$ to be complete and
compact, i.e.
without boundary, $\partial M=\emptyset$, and the metric $g$ to be positive
definite.

Let $V(M)$ be a smooth vector bundle over the manifold $M$,
${\rm End}\,(V)$ be the bundle of all smooth
endomorphisms of the vector bundle $V$, and
$C^\infty(M,V)$ and $C^\infty(M,{\rm End}\,(V))$ be the spaces of all 
smooth sections of the vector bundles $V$ and ${\rm End}\,(V)$.
We assume, as usual, the vector bundle $V$ to be Hermitian, i.e. 
there is a Hermitian pointwise fibre scalar product. Then the dual vector 
bundle $V^*$ is naturally identified with $V$ and a natural $L^2$ inner
product
is defined using  the invariant Riemannian volume element $d\vol(x)$ on the
manifold $M$.
The completion of $C^\infty(M,V)$ in this norm defines the Hilbert space of
square
integrable sections of the vector bundle $L^2(M,V)$.

Let, further, $\nabla$ be a connection, or covariant derivative, 
on the vector bundle $V$ which is compatible with the Hermitian metric 
on the vector bundle $V$. Denoting by $T^*M$ the cotangent bundle we define
the tensor product connection on the tensor product bundle $T^*M\otimes V$
by means of the Levi-Civita connection.
Similarly, we extend the connection $\nabla$ with the help of the Levi-Civita
connection
to $C^\infty(M,V)$-valued tensors of all orders and denote it just by $\nabla$.

Usually there is no
ambiguity and the precise meaning of the covariant derivative is always clear
from
the nature of the object it is acting on.

Let, further, ${\rm tr}_g$ denote the contraction of sections of the bundle 
$T^*M\otimes T^*M\otimes V$ with the metric on the cotangent bundle,
and $Q\in C^\infty(M,\End(V))$ be a a smooth Hermitian section of the
endomorphism bundle.
Then we define the generalized Laplacian
\be
\sq={\rm tr}_g \nabla\nabla
\ee
and a {\it Laplace type} differential operator $F:\ C^\infty(M,V)\to
C^\infty(M,V)$ by
\be
F=-\sq+Q.
\label{100}
\ee

Let $x^\mu$, $(\mu=1,2,\dots, d)$, be a system of local coordinates and
$\partial_\mu$ and $dx^\mu$ be the local coordinate frames for 
the tangent and the cotangent bundles.
We adopt the notation that the Greek indices label the 
tensor components with respect to local 
coordinate frame and range from 1 through $d$. 
Besides, a summation is always carried out over repeated indices.
Let $g_{\mu\nu}=(\partial_\mu,\partial_\nu)$ be the metric on the tangent
bundle, 
$g^{\mu\nu}=(dx^\mu,dx^\nu)$ be the metric on the cotangent bundle, 
$g=\det g_{\mu\nu}$, and ${\cal A}_\mu$ be the connection $1-$form on the
vector
bundle $V$.
Then the generalized Laplacian reads
\be
\sq=g^{\mu\nu}\nabla_\mu\nabla_\nu=
g^{-1/2}(\partial_\mu+{\cal A}_\mu)g^{1/2}g^{\mu\nu}
(\partial_\nu+{\cal A}_\nu).
\ee

It is worth noting that every second-order differential operator with a scalar

leading symbol given by the metric tensor is of Laplace type and 
can be put in the form (\ref{100}) by choosing the appropriate connection
$1$-form
${\cal A}$ and the endomorphism $Q$.

The operator $F$ has a positive leading symbol and is elliptic.
It is symmetric with respect to the natural $L^2$ inner product. 
Moreover, the operator $F$ is essentially self-adjoint, i.e. there is a unique

self-adjoint extension $\bar F$ of the operator $F$. 
However, we will not be very careful about distinguishing 
between the operator $F$ and its closure $\bar F$, and will simply say that the
operator
$F$ is elliptic and self-adjoint.

\subsection{Heat kernel of a Laplace type operator}

{}Hence, the operator $U_F(t)=\exp(-tF)$ for $t>0$ is well defined as
a {\it bounded} operator on the Hilbert space $L^2(M,V)$ 
of square integrable sections of the vector bundle $V$. 
This operator forms a one-parameter semi-group.
The kernel $U_F(t|x,x')$ of this operator  satisfies the heat equation and is
called
the heat kernel.
{}For $t>0$ the heat kernel is a smooth section of the external tensor product
of the vector bundles $V$ and $V^*$ over the tensor product manifold $M\times
M$.

In the following we are going to study the Green function and the 
heat kernel only locally (like in \cite{avr91b}), i.e.
in the neighbourhood of the diagonal of $M\times M$.
We will keep a point $x'$ of the manifold fixed and consider a
small geodesic ball, i.e. a small neighbourhood of the point $x'$: 
$B_{x'}=\{x\in M| r(x,x')<\varepsilon\}$, $r(x,x')$ being the geodesic distance
between the 
points $x$ and $x'$. 
We will take the radius of the ball sufficiently small, so that each 
point $x$ of the ball
of this neighbourhood can be connected by a unique geodesic 
with the point $x'$. This can be always
done if the size of the ball is smaller than the injectivity radius of the
manifold at $x'$,
$\varepsilon<r_{\rm inj}(x')$.

Let $\sigma(x,x')$ be the geodetic interval, also called world function, 
defined as one half the square of the length
of the geodesic connecting the points $x$ and $x'$
\be
\sigma(x,x')={1\over 2}r^2(x,x').
\ee
The first derivatives of this function with respect to $x$ and $x'$ 
define the tangent vector fields to the geodesic at the end points
$x$ and $x'$
and the determinant of the mixed second derivatives defines a so called 
Van Vleck-Morette determinant \cite{dewitt65,dewitt60,synge60}
\be
\Delta(x,x')=g^{-1/2}(x)\det(-\nabla_\mu\nabla'_{\nu'}
\sigma(x,x'))g^{-1/2}(x').
\ee
Let, finally, ${\cal P}(x,x')$ denote the paralell transport operator along the
geodesic
from the point $x'$ to the point $x$. 
It is a section of the external tensor product of the
vector bundle $V$ and $V^*$ over $M\times M$, or, in other words, it is an
endomorphism from the fiber of $V$ over $x'$ to the fiber of $V$ over $x$.
Near the diagonal of $M\times M$ all these two-point functions are smooth
single-valued 
functions of the coordinates of the points $x$ and $x'$. 
To simplify the consideration, one can assume that these functions are
{\it analytic} in the ball $B_{x'}$.

The heat kernel of a Laplace type operator is well described by factorizing out
the
semi-classical factor \cite{dewitt65,avr86b,avr91b}
\be
U_F(t|x,x')=(4\pi t)^{-d/2}\Delta^{1/2}(x,x')
\exp\left(-{1\over 2t}\sigma(x,x')\right)
\Omega_F(t|x,x').
\label{150}
\ee
The function $\Omega_F(t|x,x')$ is called the {\it transfer function} of the
operator $F$.
Obviously,
\be
U_{F+m^2}(t)=e^{-tm^2}U_F(t),\qquad
\Omega_{F+m^2}(t)=e^{-tm^2}\Omega_F(t).
\label{580}
\ee
It can be proved \cite{avr91b}
that for a positive operator $F+m^2$ the transfer function 
$\Omega_{F+m^2}(t)$ can be presented in form of an inverse Mellin transform
(we slightly change notation here in comparison to \cite{avr91b})
\be
\Omega_{F+m^2}(t)={1\over 2\pi
i}\int\limits_{c-i\infty}^{c+i\infty}dq\,t^q\Gamma(-q)
b_{F+m^2}(q),
\label{582}
\ee
where $c$ is a negative constant. 
The function $b_{F+m^2}(q)$ is an {\it entire} function of $q$, so 
that the integrand has simple poles at the positive integer points $q=k,\,
(k=0,1,\dots)$.
The function $b_{F+m^2}(q|x,x')$ satisfies a functional-differential equation 
\cite{avr91b}
\be
\left(1+{1\over q}D\right)b_{F+m^2}(q)
=L\,b_{F+m^2}(q-1),
\label{12a}
\ee
where
\be
L=\Delta^{-1/2}F\Delta^{1/2}+m^2=-\Delta^{-1/2}\sq\Delta^{1/2}+Q+m^2,
\ee 
\be
D=\sigma^\mu\nabla_\mu, \qquad\sigma_\mu=\nabla_\mu\sigma,
\ee
and the initial condition
\be
b_{F+m^2}(0|x,x')={\cal P}(x,x').
\ee
This equation (together with the condition of analyticity and 
some asymptotic condition at
$q\to {\rm const}\pm i\infty$) enables one to compute $b_{F+m^2}(q)$
if one fixes its initial value at some arbitrary point $q=q_0$. 
(For more details, see \cite{avr91b}).
The  initial condition at the origin produces the values of $b_{F+m^2}(k)$ at
positive integer points, the initial condition at the point $q=1/2$ would give
the values of $b_{F+m^2}(q)$ at all half-integer positive points 
$b_{F+m^2}(k+1/2)$, etc.  Differentiating the eq. (\ref{12a}) with respect
to $q$ one obtains another recursion 
\be
\left(1+{1\over q}D\right)b'_{F+m^2}(q)
=L\,b'_{F+m^2}(q-1)+{1\over q^2}Db_{F+m^2}(q),
\label{12b}
\ee
where
\be
b'_{F+m^2}(q)={\partial\over\partial q}b_{F+m^2}(q),
\ee
which enables one to compute the derivatives of the function $b_{F+m^2}(q)$
at positive integer points if one fixes its value $b'_{F+m^2}(0)$.

Moving the contour of integration to the right and taking into account that the
residues
of the gamma-function $\Gamma(-q)$ at the points $q=k$ equal $(-1)^k/k!$ 
one obtains \cite{avr91b}
\be
\Omega_{F+m^2}(t)=\sum_{k=0}^{n-1}{(-t)^k\over k!}a_k(F+m^2)
+{1\over 2\pi i}\int\limits_{c_n-i\infty}^{c_n+i\infty}dq\,t^q\Gamma(-q)
b_{F+m^2}(q),
\label{596}
\ee
where
\be
a_k(F+m^2)\equiv b_{F+m^2}(k)= 
\left(-{\partial\over\partial t}\right)^k\Omega_{F+m^2}(t)\Big|_{t=0},
\label{590}
\ee
$n$ is an arbitrary positive integer, $n\ge 1$, and $c_n$ ranges 
in the interval $n-1<c_n<n$.
The coefficients $a_k(F+m^2)$ are the famous
Hadamard-Minakshisundaram-De Witt-Seeley (HMDS)-coefficients
\cite{hadamard23,minakshisundaram53,dewitt65,seeley67b} to the operator
$F+m^2$.
They can be computed in form of covariant Taylor series
from the recursion relations that are obtained from the equation (\ref{12a}) 
(by putting $q\to k$) with the initial condition $a_0(F+m^2)={\cal P}$. 
The diagonal values of the HMDS-coefficients 
are known in general case up to $a_4$ 
\cite{gilkey75b,avr86b,avr90b,avr91b,amsterdamski89}.
In some particular cases, e.g. in flat space, there are results for
higher-order
coefficients. For a review of the methods for calculation of HMDS-coefficients

and further references see 
\cite{avr-s96,schimming93}.

Taking into account eq. (\ref{580}) we find from (\ref{590})
\be
a_k(F+m^2)=\sum_{n=0}^k {k\choose n}m^{2(k-n)}a_n(F).
\label{15}
\ee
Moreover, one can obtain an asymptotic expansion of the function
$b_{F+m^2}(q)$
as $m\to\infty$ \cite{avr91b}
\be
b_{F+m^2}(q)\sim\sum_{n\ge 0} {\Gamma(q+1)\over n!
\Gamma(q-n+1)}\,m^{2(q-n)}a_n(F).
\label{15c}
\ee

{}From (\ref{596}) there follows an asymptotic expansion of the transfer
function
as $t\to 0$ (called Schwinger-De Witt expansion in the physical literature)
\cite{dewitt65,gilkey84,avr86b,avr91b,barvinsky85}
\be
\Omega_{F+m^2}(t)\sim \sum\limits_{k\ge 0}{(-t)^k\over k!}a_k(F+m^2)=
e^{-tm^2}\sum\limits_{k\ge 0}{(-t)^k\over k!}a_k(F).
\label{600}
\ee

{}Using this equation one obtains from (\ref{150})
the well known formula for the asymptotic expansion of the 
functional trace of the heat kernel
\be
\Tr_{L^2}\exp(-tF)\sim (4\pi)^{-d/2} t^{-d/2}\sum\limits_{k\ge 0}{(-t)^k\over
k!}A_k(F),
\label{600c}
\ee
where
\be
A_k(F)=\Tr_{L^2}a_k(F)=\int\limits_M d\vol(x)\,\tr_V a_k(F|x,x),
\ee
with $\tr_V$ being the fiber trace.

\subsection{Green function of a Laplace type operator}

The Laplace type operator $F$ is a self-adjoint elliptic operator with a real 
spectrum bounded from below. Thus for sufficiently large $m$ the operator 
$(F+m^2)$ is positive and its Green operator, or the resolvent operator, 
$G_{F+m^2}=(F+m^2)^{-1}$ is a bounded operator. It can be defined by
\be
G_{F+m^2}=\int\limits_0^\infty dt\,e^{-tm^2}\exp(-tF).
\label{200}
\ee
The Green function of the operator $(F+m^2)$ is obtained by the kernel form of

this equation 
\be
G_{F+m^2}=(4\pi)^{-d/2}\Delta^{1/2}
\int\limits_0^\infty dt\,t^{-d/2}
\exp\left(-tm^2-{\sigma\over 2t}\right)\Omega_F(t).
\label{300}
\ee

This integral converges at the infinity for sufficiently large $m$. It also
converges at $t=0$ outside the diagonal, i.e. for $\sigma\ne 0$.
By using the ansatz (\ref{150}) and (\ref{582}) for the heat kernel 
one obtains a corresponding ansatz for the Green function
\be
G_{F+m^2}=(4\pi)^{-d/2}\Delta^{1/2}
{1\over 2\pi i}\int\limits_{c-i\infty}^{c+i\infty}dq\,\left({\sigma\over
2}\right)^{q+1-d/2}
\Gamma(-q)\Gamma(-q-1+d/2)b_{F+m^2}(q),
\label{597a}
\ee
where $c<-1/2$.

This ansatz is especially useful for studying the singularities of the Green
function, or more
general, for constructing the Green function as a power series in $\sigma$.
The integrand in (\ref{597a}) is again a meromorphic function. However, 
contrary to (\ref{582}), we have now a more complicated structure of the
poles.
There are always poles at the points $q=k$ and $q=k-1+d/2$,\
$(k=0,1,2,\dots)$.
Here one has to distinguish between odd and even dimensions.
In odd dimensions, the poles are 
at the points $q=k$ and $q=k+[d/2]-1/2$ and are {\it simple}, whereas in even
dimension there are simple poles at $q=0,1,2,\dots,d/2-2$ and {\it double}
poles at the points $q=k+d/2-1$.

Moving the contour of integration in (\ref{597a}) to the right one can obtain 
an expansion of the Green function
in powers of $\sigma$ (Hadamard series).
Generally, we obtain
\be
G_{F+m^2}=G^{\rm sing}_{F+m^2}
+G^{\rm non-anal}_{F+m^2}
+G^{\rm reg}_{F+m^2}.
\ee
Here $G^{\rm sing}_{F+m^2}$ is the singular part which is polynomial
in the inverse powers of $\sqrt\sigma$
\be
G^{\rm sing}_{F+m^2}=
(4\pi)^{-d/2}\Delta^{1/2}
\sum_{k=0}^{[(d+1)/2]-2}{(-1)^k\over
k!}\Gamma(d/2-k-1)\left({2\over\sigma}\right)^{d/2-k-1}
a_k(F+m^2),
\ee
For the rest we get in {\it odd} dimensions
\bea
&&G^{\rm non-anal}_{F+m^2}
+G^{\rm reg}_{F+m^2}
\nonumber\\
&&=
(-1)^{d-1\over 2}(4\pi)^{-{d\over 2}}\Delta^{1\over 2}
\sum_{k=0}^{n-(d+1)/2}
{\pi\over \Gamma\left(k+{d+1\over 2}\right)\Gamma\left(k+{3\over 2}\right)}
\left({\sigma\over 2}\right)^{k+1/2}
a_{k+{d-1\over 2}}(F+m^2)
\nonumber\\
&&
+(-1)^{(d+1)/2}(4\pi)^{-d/2}\Delta^{1/2}\sum_{k=0}^{n-(d+1)/2}{\pi\over k!
\Gamma(k+d/2)}\left({\sigma\over 2}\right)^{k}
b_{F+m^2}(k-1+d/2)
\nonumber\\[11pt]
&&+
(4\pi)^{-d/2}\Delta^{1/2}
{1\over 2\pi i}\int\limits_{c_n-i\infty}^{c_n+i\infty}dq\,\left({\sigma\over
2}\right)^{q+1-d/2}
\Gamma(-q)\Gamma(-q-1+d/2)b_{F+m^2}(q),
\nonumber\\
\label{597b}
\eea
where $n-1<c_n<n-1/2$ and $n>(d-1)/2$.
Thus, by putting $n\to\infty$ we recover herefrom the Hadamard 
power series in $\sigma$ for {\it odd} $d=1,3,5,\dots$
\be
G^{\rm non-anal}_{F+m^2}\sim
(-1)^{d-1\over 2}(4\pi)^{-{d\over 2}}\Delta^{1\over 2}
\sum_{k=0}^{\infty}
{\pi\over \Gamma\left(k+{d+1\over 2}\right)\Gamma\left(k+{3\over 2}\right)}
\left({\sigma\over 2}\right)^{k+1/2}
a_{k+{d-1\over 2}}(F+m^2)
\ee
\be
G^{\rm reg}_{F+m^2}\sim
(-1)^{(d+1)/2}(4\pi)^{-d/2}\Delta^{1/2}\sum_{k=0}^{\infty}{\pi\over k!
\Gamma(k+d/2)}\left({\sigma\over 2}\right)^{k}
b_{F+m^2}(k-1+d/2).
\ee

In {\it even} dimensions, the point is more subtle due to the presence of
double poles.
Moving the contour in (\ref{597a}) to the right and calculating the
contribution of the 
residues at the simple and double poles we obtain
\bea
&&G^{\rm non-anal}_{F+m^2}
+G^{\rm reg}_{F+m^2}\nonumber\\
&&=
(-1)^{d/2-1}(4\pi)^{-d/2}\Delta^{1/2}\log\left(\mu^2\sigma\over 2\right)
\sum_{k=0}^{n-1}{1\over k!
\Gamma(k+d/2)}\left({\sigma\over 2}\right)^{k}
a_{k-1+d/2}(F+m^2)
\nonumber\\[11pt]
&&
+(-1)^{d/2-1}(4\pi)^{-d/2}\Delta^{1/2}\sum_{k=0}^{n-1}
{1\over k!\Gamma(k+d/2)}\left({\sigma\over 2}\right)^{k}
\nonumber\\[11pt]
&&\times\left\{
b'_{F+m^2}(k-1+d/2)
-\left[\log\mu^2+\Psi(k+1)+\Psi(k+d/2)\right]
a_{k-1+{d\over 2}}(F+m^2)\right\}
\nonumber\\[11pt]
&&
+(4\pi)^{-d/2}\Delta^{1/2}
{1\over 2\pi i}\int\limits_{c_n-i\infty}^{c_n+i\infty}dq\,\left({\sigma\over
2}\right)^{q+1-d/2}
\Gamma(-q)\Gamma(-q-1+d/2)b_{F+m^2}(q),
\nonumber\\
\label{597}
\eea
where $\mu$ is an arbitrary mass parameter introduced to preserve dimensions,
$n-1<c_n<n$ and $\Psi(z)=(d/dz)\log\,\Gamma(z)$.
If we let $n\to\infty$ we obtain the Hadamard expansion of the Green function
for {\it even} $d=2,4,\dots$
\be
G^{\rm non-anal}_{F+m^2}\sim
(-1)^{d/2-1}(4\pi)^{-d/2}\Delta^{1/2}\log\left(\mu^2\sigma\over 2\right)
\sum_{k=0}^{\infty}
{1\over k!\Gamma(k+d/2)}\left({\sigma\over 2}\right)^{k}
a_{k-1+d/2}(F+m^2)
\label{597f}
\ee
\bea
G^{\rm reg}_{F+m^2}&\sim&
(-1)^{d/2-1}(4\pi)^{-d/2}\Delta^{1/2}\sum_{k=0}^{\infty}
{1\over k!\Gamma(k+d/2)}\left({\sigma\over 2}\right)^{k}
\Biggl\{b'_{F+m^2}(k-1+d/2)
\nonumber\\[11pt]
&&
-\left[\log\mu^2+\Psi(k+1)+\Psi(k+d/2)\right]
a_{k-1+{d\over 2}}(F+m^2)\Biggr\}
\nonumber\\
\eea

Note that the singular part
(which is a polynomial in inverse powers of $\sqrt\sigma$)
and
the {\it non-analytical} parts
(proportional to $\sqrt\sigma$ and $\log\sigma$)
are expressed {\it uniquely only in terms of the local HMDS-coefficients} 
$a_k(F+m^2)$, whereas the regular analytical part 
contains the values of the function $b_{F+m^2}(q)$ at half-integer
positive points and the derivatives of the function
$b_{F+m^2}(q)$ at integer positive points,
which are {\it not} expressible in terms of the local information.
These objects are global and cannot be expressed
further in terms of the local HMDS-coefficients.
However, they can be computed from the eqs. (\ref{12a}) and (\ref{12b})
almost by the same algorithm as the HMDS-coefficients
in terms of the value of the function $b(q)$ at some fixed $q_0$
(see \cite{avr91b}).

The regular part of the Green function has a well defined diagonal value 
and the functional trace. It reads in odd dimensions $(d=1,3,5,\dots)$:
\bea
\Tr_{L^2}\,G^{\rm reg}_{F+m^2}
&=&(-1)^{(d+1)/2}(4\pi)^{-d/2}{\pi\over \Gamma(d/2)}
B_{F+m^2}(d/2-1)
\nonumber\\
&\stackrel{m\to\infty}{\sim} &(-1)^{(d+1)/2}(4\pi)^{-d/2}\pi
\sum_{k\ge 0}{m^{d-2-2k}\over k!\Gamma(d/2-k)}A_k(F)
\label{25}
\eea
and in even dimensions $(d=2,4,6,\dots)$
\bea
\Tr_{L^2}\,G^{\rm reg}_{F+m^2}&=&
(-1)^{d/2-1}{(4\pi)^{-d/2}\over \Gamma(d/2)}\Biggl\{
B'_{F+m^2}(d/2-1)
\nonumber\\[11pt]&&
-\left[\log\,\mu^2+\Psi(d/2)-{\CC }\right]
A_{d/2-1}(F+m^2)\Biggr\}
\nonumber\\
&\stackrel{m\to\infty}{\sim}&
(-1)^{d/2-1}(4\pi)^{-d/2}
\Biggl\{
\sum_{k=0}^{d/2-1}{m^{d-2-2k}\over k!\Gamma(d/2-k)}
[\CC-\Psi(d/2-k)+\log\,{m^2\over \mu^2}]A_k(F)
\nonumber\\
&&
+\sum_{k\ge d/2}{(-1)^{k-d/2}\over k!}m^{d-2-2k}\Gamma(k+1-d/2)
A_k(F)\Biggr\},
\label{33}
\eea
where
\be
B_{F+m^2}(q)=\Tr_{L^2}b_{F+m^2}(q)=\int\limits_Md\vol(x)\tr_V\,b_{F+m^2}(x,x),
\ee
and ${\CC }=-\Psi(1)=0.577\dots$ is the Euler's constant.

Thus, we see that
\begin{itemize}
\item[i)]
all the singularities of the Green function and the non-analytical parts
thereof
(proportional to $\sqrt\sigma$ in odd dimensions and to $\log\sigma$ in even
dimensions)
are determined by the HMDS-coefficients $a_k(F)$;
\item[ii)]
there are no power singularities, i.e. $G^{\rm sing}_{F+m^2}=0$,
in lower dimensions $d=1,2$;
\item[iii)]
there is no logarithmic singularity (more generally, no logarithmic part at
all) in odd dimensions;
\item[iv)]
the regular part depends on the values of the function $b_{F+m^2}(q)$ at
half-integer
points and its derivative at integer points and is a global object that cannot
be reduced
to purely local information like the HMDS-coefficients.
\end{itemize}

The logarithmic part of the Green function
is very important.
On the one hand it determines, as usual, 
the renormalization properties of the
regular part of the Green function, i.e. the derivative
$\mu(\partial/\partial\mu)G^{\rm reg}_{F+m^2}$.
In particular,
\be
\mu{\partial\over\partial\mu}\Tr_{L^2}G^{\rm reg}_{F+m^2}
=\left\{
\begin{array}{ll}
0 & {\rm for\ odd\  } d\\
{\displaystyle
{(4\pi)^{-d/2}\over\Gamma(d/2)}A_{d/2-1}(F+m^2)} & {\rm for\ even\ } d.
\end{array}
\right.
\ee

On the other hand, it is of crucial importance
in studying the Huygens principle. 
Namely, the absence of the logarithmic
part of the Green function is a necessary and sufficient
condition for the validity of the Huygens principle for hyperbolic operators 
\cite{schimming78,schimming82,schimming90}. 
The HMDS-coefficients and, therefore, the logarithmic part of the Green
function are defined
for the hyperbolic operators just by analytic continuation from the elliptic
case. Thus,
the condition of the validity of Huygens principle reads
\be
\sum_{k=0}^{\infty}
{\Gamma(d/2)\over k!\Gamma(k+d/2)}\left({\sigma\over 2}\right)^{k}
a_{k-1+d/2}(F+m^2)=0,
\label{597c}
\ee
or, by using (\ref{15}),
\be
\sum_{k=0}^{\infty}\sum_{n=0}^{k-1+d/2}
{\Gamma(d/2)\over k!n!\Gamma(k-n+d/2)}\left({\sigma\over 2}\right)^{k}
m^{2(k-n)}a_{n}(F)=0,
\label{597d}
\ee
By expanding this equation in covariant Taylor series using the methods of 
\cite{avr91b} one can obtain an infinite set of local conditions for validity
of the Huygens
principle.

\be
\sum_{k=0}^{[n/2]}
{n!\Gamma(d/2)\over 4^k k!(n-2k)!\Gamma(k+d/2)}
(\vee^k g)\vee<n-2k|a_{k-1+d/2}(F+m^2)>=0,
\label{597g}
\ee
where $\vee$ is the exterior symmetric tensor product,
$g$ is the metric tensor on the tangent bundle and $<n|a_k>$ denotes the 
diagonal value of the symmetrized covariant derivative of $n$-th order
\cite{avr91b}.
More explicitly,
\bea
&&[a_{d/2-1}(F+m^2)]=0,\\
&&[\nabla_\mu a_{d/2-1}(F+m^2)]=0,\\
&&[\nabla_{(\mu}\nabla_{\nu)}a_{d/2-1}(F+m^2)]
+{1\over 2d}g_{\mu\nu}[a_{d/2}(F+m^2)]=0,\\
&&\dots\nonumber
\eea
where the square brackets denote the diagonal value of
two-point quantities: $[f(x,x')]=f(x,x)$.

\section{HIGHER ORDER OPERATORS}

Let us now describe in detail the class of higher order operators we are going
to study.
We consider again a Riemannian manifold $(M,g)$ and a vector bundle $V(M)$
with a connection $\nabla$.
Let ${\cal A}_{(i)}$, $(i=1,\dots,N)$ be a set of different 
smooth sections of the vector bundle $T^*M\otimes\End(V)$.
They define a set of {\it different} connections
\be
\nabla_{(i)}=\nabla+{\cal A}_{(i)}
\ee
on the vector bundle $V(M)$, and therefore, a set of different generalized
Laplacians 
\bea
\sq_{i}&=&\tr_g\nabla_{(i)}\nabla_{(i)}=g^{\mu\nu}(\nabla_\mu+{\cal
A}_{(i)\,\mu})
(\nabla_\nu+{\cal A}_{(i)\,\nu})
\nonumber\\[11pt]
&=&\sq+g^{\mu\nu}({\cal A}_{(i)\mu}\nabla_\nu
+\nabla_\nu{\cal A}_{(i)\mu})
+g^{\mu\nu}{\cal A}_{(i)\mu}{\cal A}_{(i)\nu}.
\eea
Let, further, $Q_i$ be a set of different smooth sections of the vector bundle
$\End(V)$. 
Then, we define a set of {\it different} Laplace type operators
\be
F_i=-\sq_i+Q_i
\ee
and a higher order operator of a special form
\be
H=(F_N+m^2)\cdots (F_2+m^2)(F_1+m^2).
\label{700}
\ee

\subsection{Algebraical reduction to Laplace type operators}

We are going to study the Green function $G_{H-\lambda}(x,x')$ of the 
higher order operator $H-\lambda$, where $\lambda$ is an arbitrary sufficiently

large negative constant.
To do this, we show, first, that the Green operator $G_{H-\lambda}
=(H-\lambda)^{-1}$
can be reduced to the Green operator of an auxiliary Laplace type operator.
This will mean that the Green function
of the higher order differential operator will be reduced to the Green function
of a {\it second-order} differential operator of Laplace type, which is
described in previous section.

\begin{theorem}
Let $F_i$, $(i=1,\dots,N)$, be some positive elliptic operators and $H$ be an
operator defined by
\be
H=(F_N+m^2)\cdots (F_2+m^2)(F_1+m^2),
\label{701}
\ee
where $m$ is a sufficiently large constant.
Let $I_{N-1}$ be the unit $(N-1)\times (N-1)$ matrix
and $O_{N-1}$ be the square $(N-1)\times (N-1)$ null matrix.
Let $P$ and $\Pi$ be square $N\times N$ matrices defined by the following
block form
\be
P=\left(
\begin{array}{cc}
0&I_{N-1}\\
\lambda&0
\end{array}
\right),
\qquad
\Pi={\partial\over\partial\lambda}P
=\left(
\begin{array}{cc}
0&O_{N-1}\\
1&0
\end{array}
\right),
\label{17a}
\ee
where $\lambda$ is a constant,
and let $\widetilde F=F_1\oplus\cdots\oplus F_N$ be a $N\times N$
diagonal matrix formed by the operators $F_i$
\be
\widetilde F=\left(
\begin{array}{ccc}
F_1&\dots&0\\
\vdots&\ddots&\vdots\\
0&\dots&F_N
\end{array}
\right)
\ee
Let ${\cal F}$ be an operator defined by
\be
{\cal F}=\widetilde F-P.
\label{800}
\ee
Let $G_{H-\lambda}$ and $G_{{\cal F}+m^2}$ be the Green operators
of the operators $(H-\lambda)$ and $({\cal F}+m^2)$ 
satisfying the equations
\be
(H-\lambda)G_{H-\lambda}=1
\label{810}
\ee
and
\be
({\cal F}+m^2)G_{{\cal F}+m^2}=1.
\ee
Then there holds
\be
G_{H-\lambda}={\rm tr}\,\Pi\,
G_{{\cal F}+m^2},
\label{16}
\ee
\be
{\partial\over\partial m^2}G_{H-\lambda}
=-{\partial\over\partial\lambda}{\rm tr}\,G_{{\cal F}+m^2},
\label{50}
\ee
\end{theorem}

\paragraph{Proof.}\
Let us introduce some auxiliary operators
\bea
Z_1&=&G_{H-\lambda}\nonumber\\
Z_2&=&(F_1+m^2)Z_1\nonumber\\
&\vdots&
\label{500}\\
Z_N&=&(F_{N-1}+m^2)Z_{N-1}
\nonumber.
\eea
Then the equation (\ref{810})
can be rewritten as
\be
(F_N+m^2)Z_N-\lambda G_{H-\lambda}=1.
\ee
Let us collect the operators $Z_i$ in a column-vector
\be
Z=\left(
\begin{array}{c}
Z_1\\
\vdots\\
Z_N
\end{array}
\right)
\ee
The Green operator $G_{H-\lambda}$ is the first component of the vector $Z$ 
and can be obtained by multiplying with the row-vector
$\Pi^{\dag}_1=(1,0,\dots,0)$ 
\be
G_{H-\lambda}=\Pi^{\dag}_1 Z.
\label{14a}
\ee
Further, let us define another column-vector 
\be
\Pi_N=\left(
\begin{array}{c}
0\\
\vdots\\
0\\
1
\end{array}
\right),
\ee
so that
\be
\Pi=\Pi_N\otimes \Pi_1^{\dag}.
\label{750}
\ee
It is not difficlut to show that the equations (\ref{500})
can be rewritten in a matrix form
\be
({\cal F}+m^2)Z=\Pi_N.
\label{14}
\ee
Using the Green operator $G_{{\cal F}+m^2}$ of the operator
$({\cal F}+m^2)$
we obtain
\be
Z=G_{{\cal F}+m^2}\Pi_N.
\ee
Finally, from (\ref{14a}) by taking into account (\ref{750}) we get the Green
operator $G_{H-\lambda}$
\be
G_{H-\lambda}=\Pi^{\dag}_1\,G_{{\cal F}+m^2}\Pi_N
={\rm tr}\,\Pi\,
G_{{\cal F}+m^2},
\label{16a}
\ee
which proves the eq. (\ref{16}).
To prove eq. (\ref{50}) we note that
\bea
{\partial\over\partial\lambda}{\rm tr}\,G_{{\cal F}+m^2}
&=&-\tr\,G_{{\cal F}+m^2}
{\partial {\cal F}\over\partial\lambda}
G_{{\cal F}+m^2}
\nonumber\\
&=&-\tr\,\Pi\,G^2_{{\cal F}+m^2}.
\label{60a}
\eea
Further, from the definition of the Green operator we have, obviosly,
\be
{\partial\over\partial m^2}\tr\,\Pi\,G_{{\cal F}+m^2}
=-\tr\,\Pi\,G^2_{{\cal F}+m^2}.
\label{60b}
\ee
Comparing (\ref{60a}) and (\ref{60b}) and using the eq. (\ref{16a}) we
convince ourselves that eq. (\ref{50}) is correct too.
\medskip

Let us make some remarks.
First, ${\rm `tr'}$ denotes here
the usual matrix trace and has nothing to do with the trace over
 the bundle indices which are left intact.
Second, the operator $\widetilde F$ has, obviously, the form
\be
\widetilde F=-\widetilde{\sq}+\widetilde Q,
\ee
where $\widetilde{\sq}=\sq_1\oplus\cdots\oplus\sq_N$ and 
$\widetilde Q=Q_1\oplus\cdots\oplus Q_N$ have the same diagonal structure.
Obviously
\be
 \widetilde{\sq}=\tr_g\widetilde{\nabla}\widetilde{\nabla}
\ee
where $\widetilde{\nabla}={\nabla}_1\oplus\cdots\oplus
{\nabla}_N $ is a new connection which also has the same diagonal
structure.
Therefore, the operator ${\cal F}$ reads
\be
{\cal F}=-\widetilde{\sq}+\widetilde Q-P.
\ee
Remember that the matrix $P$ is constant. 
Thus, both the operator $\widetilde F$ and ${\cal F}$ are
{\it second-order} elliptic Laplace type differential operators.
However, the operator ${\cal F}$
is {\it not self-adjoint} because the matrix $P$ is
not symmetric. In spite of this, the operator
$({\cal F}+m^2)$ for sufficiently large $m$ is non-degenerate, i.e.
there exists a well defined Green operator $G_{{\cal F}+m^2}$, which
justifies made assumptions.

\subsubsection{Properties of the matrix $P$}

The matrix $P$ plays the
role of the matrix term and is of great importance in
further considerations. That is why, we state below some important properties
of it.

First of all, we note that the matrix $\Pi$ is nilpotent
\be
\Pi^2=0
\ee
and, of course, degenerate.
The matrix $P$ is not degenerate. 
The matrix $P$ as well as its powers belong to the class of two-diagonal
matrices. We compute first the powers of this matrix.
\be
P=\left(
\begin{array}{cc}
0&I_{N-1}\\
\lambda &0
\end{array}
\right),\quad
P^2=\left(
\begin{array}{cc}
0&I_{N-2}\\
\lambda I_2&0
\end{array}
\right),\
\cdots,\
P^{N-1}=\left(
\begin{array}{cc}
0&1\\
\lambda I_{N-1}&0
\end{array}
\right),
\label{59}
\ee
\be
P^{N}=\lambda I_{N}.
\label{60}
\ee
It is also not difficult to show that
\be
\det\left(z-P\right)=z^N-\lambda.
\ee
Therefore, the eigenvalues $p_j$ of the matrix $P$ are just the
$N$-th roots of $\lambda$
\be
p_j=(-\lambda)^{1/N}e^{i\,\pi(2j-1)/N},
\qquad j=1,\dots,N,
\ee
where $|{\rm arg}\,(-\lambda)|<\pi$. The roots $p_j$
lie on the circle $|z|=|\lambda|^{1/N}$.

The higher powers of the matrix $P$ are expressed in terms of the lower order
powers
(\ref{59}) and (\ref{60})
\be
P^{iN+j}=\lambda^i P^{j},
\label{63}
\ee
where $j=0,1,2,\dots,N-1$ and $i=0,1,2,\dots$.
Using these formulas we find that for
any function $f(z)$ that is analytic at the points $p_j(\lambda)$ there holds
\bea
f(P)&=&{1\over 2\pi i}\int\limits_{C_P} 
dz\, f(z)\,(z-P)^{-1}\nonumber\\
&=&
\sum_{j=0}^{N-1}\beta_{N-1-j}(\lambda;f)P^j
\label{65}
\eea
where
\be
\beta_n(\lambda;f)=
{1\over 2\pi i}\int\limits_{C_P} 
dz\,{z^{n}\over z^N-\lambda}f(z)
\label{70a}
\ee
and the contour $C_P$ goes counter-clockwise around the points 
$p_j$ in such
a way that it does not contain the singularities
of the function $f(z)$ --- inside the contour $C_P$ 
there must be no other singularities
except for the points $p_j$.
Evidently, the coefficients $\beta_n(\lambda;f)$ possess the following
`periodicity' property
\be
\beta_{n+kN}(\lambda;f)=\lambda^k\beta_n(\lambda;f),
\qquad k=0,1,2,\dots,
\ee
so that there are essentially only $N$ independent coefficients
$\beta_0, \beta_1, \dots, \beta_{N-1}$. 

The eqs. (\ref{65}) and (\ref{67}) determine any function of the matrix $P$.
In particular, using the traces of the powers of the matrix $P$
\be
\tr\,P^{n}
=\left\{
\begin{array}{ll}
N\lambda^i & {\rm for }\ n=iN,\ i=0,1,2,\dots,\\
0 & {\rm otherwise}
\end{array}
\right.
\label{72a}
\ee
we can compute the trace of any function of $P$
\be
\tr\,f(P)=N\beta_{N-1}(\lambda;f).
\ee
More generally,
\be
\tr\,P^n\,f(P)=N
\beta_{N-1+n}(\lambda;f).
\ee
Herefrom we obtain the coefficients $\beta_n(\lambda;f)$
\bea
\beta_n(\lambda;f)&=&{1\over N\lambda}\sum_{j=1}^N p_j^{n+1}f(p_j)
\nonumber\\
&=&-{1\over N}(-\lambda)^{(n-N+1)/N}
\sum_{j=1}^N e^{i\,\pi(2j-1)(n+1)/N}f(p_j)
\label{67}
\eea
{}From (\ref{72a}), it follows, in particular,
\be
\sum_{j=1}^N p^{n}_j=0
\qquad {\rm for }\ n\ne 0,\pm N,\pm 2N,\dots.
\ee

Moreover, using the definition 
(\ref{17a}) we calculate the traces of the product of the powers
of the matrix $P$ with the matrix $\Pi$
\be
\tr\,\Pi\,P^{n}=
{1\over (n+1)}{\partial\over\partial\lambda}\tr\,P^{n+1}=
\left\{
\begin{array}{ll}
\lambda^i & {\rm for }\ n=iN+N-1,\ i=0,1,2,\dots,\\
0 & {\rm otherwise}
\end{array}
\right..
\label{70}
\ee
By using the equation
\be
{\partial\over\partial\lambda}\tr\,f(P)=\tr\,{\partial P\over\partial\lambda}
\,f'(P)
\label{78c}
\ee
where $f'(z)=(\partial/\partial z)f(z)$,
and the eq. (\ref{17a}) or using directly the eq. (\ref{70}) we calculate 
the trace of the product of the function $f(P)$ with the matrix $\Pi$
\be
\tr\,\Pi\,f(P)=\beta_0(\lambda;f)
=-{1\over N}(-\lambda)^{-(N-1)/N}
\sum_{j=1}^N e^{i\,\pi(2j-1)/N}f(p_j).
\ee
More generally, one can obtain also
\be
\tr\,\Pi(z-P)^{-1}
=-{\partial\over\partial\lambda}\log\,\det(z-P)={1\over z^N-\lambda}
\ee
as well as
\be
\tr\,\Pi\,e^{tP}
={1\over 2\pi i}\int\limits_{C_P} dz {e^{tz}\over z^N-\lambda}
=-{1\over N}(-\lambda)^{-(N-1)/N}
\sum_{j=1}^N e^{i\,\pi(2j-1)/N}e^{tp_j}.
\label{80}
\ee

In the next section we will need also the following lemma,
which  is actually a generalization of the eq. (\ref{72a}).

\begin{lemma}
Let  $P$ and $\Pi$ be the $N\times N$ matrices defined by (\ref{17a}) and
$B_1,\dots,B_{N-2}$ be arbitrary diagonal matrices.
Then for any $k\le N-2$ and arbitrary non-negative integers 
$n_1,\dots,n_k$ satisfying the condition $n_1+\cdots+n_k\le N-2$ there holds

\be
\tr\,\Pi\,B_1P^{n_1}B_2P^{n_2}\cdots B_{k}P^{n_k}=0.
\label{91f}
\ee
\end{lemma}

\subsubsection{Properties of the symbol $\sigma({\cal F})$}
Let us consider now the symbol $\sigma({\cal F})$ of the operator
${\cal F}$. It has the form
\be
\sigma({\cal F})=\sigma(\widetilde{F})-P,
\ee
where $\sigma(\widetilde{F})$ is a diagonal matrix.

That is why let us consider the following matrix
\be
M\equiv\tilde\mu-P,
\label{84c}
\ee
where 
$\tilde\mu=\mu_1\oplus\cdots\oplus\mu_N$ is a diagonal
matrix with some constants $\mu_j$ on the diagonal. 
Note that the matrices $\tilde\mu$ and $P$ do not commute
with each other $[\tilde\mu,P]\ne 0$.

We are interested in the eigenvalues $q_j$ of the matrix $M$.
It is not difficult to calculate
\be
\tr M^n
=\tr{\tilde\mu}^n
=\sum_{j=1}^N\mu_j^n,
\qquad (n=1,2,\dots,N-1)
\label{87c}
\ee
\be
\tr M^N
=\sum_{j=1}^N\mu_j^N+(-1)^NN\lambda,
\label{88c}
\ee
\be
\det\left(z-M\right)
=\prod_{j=1}^N(z-\mu_j)-(-1)^N\lambda.
\ee
Thus the eigenvalues $q_j$ 
of the matrix $M$
are determined either from one algebraical equation of $N$-th order
\be
(\mu_1-z)\cdots(\mu_N-z)-\lambda=0
\label{83a}
\ee
or from the system of $N$ equations
\bea
q_1+\cdots+q_N&=&\sum_{j=1}^N\mu_j
\label{91c}\\
&\cdots&\nonumber\\
q^{N-1}_1+\cdots+q^{N-1}_N&=&\sum_{j=1}^N\mu_j^{N-1}\\
q^N_1+\cdots+q^N_N&=&\sum_{j=1}^N\mu_j^N+(-1)^NN\lambda.
\label{94c}
\eea
This enables one to calculate the traces
\be
\tr\,f(M)=\sum_{j=1}^Nf(q_j).
\ee
One can get also analogous to (\ref{70a}) integral representation
\be
\tr\,f(M)=
{1\over 2\pi i}\int\limits_{C_M}dz\,f(z)
{A(z,\mu,\lambda)\over (z-\mu_1)\cdots(z-\mu_N)-(-1)^N\lambda}
\label{70b}
\ee
where $C_M$ is a counter-clockwise contour around the spectrum of the 
matrix $M$ and
\be
A(z,\mu,\lambda)
=\sum_{j=1}^N\prod_{k=1;\,k\ne j}^{N}(z-q_k)
\ee
is a polynomial of order $(N-1)$.

Further, by using the equation
\be
{\partial\over\partial\lambda}\tr\,f(M)
=\tr\,{\partial M\over\partial\lambda}
\,f'(M),
\label{78d}
\ee
where $f'(z)=(\partial/\partial z)f(z)$,
and the obvious equation
\be
{\partial\over\partial\lambda}M=-\Pi,
\ee
which follows from the definition of the matrix $M$ and the eq. (\ref{17a}),
we calculate 
the trace of the product of the function $f'(M)$ with the matrix $\Pi$
\be
\tr\,\Pi\,f'(M)
=-{\partial\over\partial\lambda}\tr\,f(M).
\ee
Using the eqs. (\ref{87c}) and (\ref{88c}) we get, in particular
\bea
\tr\,\Pi\,M^n
&=&-{1\over n+1}{\partial\over\partial\lambda}\tr\,M^{n+1}
=0
\qquad {\rm for}\ n=0,1,\dots,N-2
\label{101c}\\
\tr\,\Pi\,M^{N-1}
&=&-{1\over N}{\partial\over\partial\lambda}\tr\,M^{N}
=(-1)^{N-1}.
\label{102c}
\eea
Note that (\ref{101c}) is a particular case of the eq. (\ref{91f}).
Further, from (\ref{91c}) -- (\ref{94c}) one obtains
\bea
\sum_{j=1}^N 
{\partial q_j\over\partial\lambda}
q^n_j&=&0
\qquad {\rm for}\ n=0,1,2,\dots,N-2.
\\
\sum_{j=1}^N 
{\partial q_j\over\partial\lambda}
q^{N-1}_j&=&(-1)^{N}.
\eea
Further, we get also
\be
\tr\,\Pi\,(z-M)^{-1}
={\partial\over\partial\lambda}\log\,\det(z-M)
={(-1)^{N-1}\over (z-\mu_1)\cdots(z-\mu_N)-(-1)^N\lambda},
\ee
\be
\tr\,\Pi\,M^{-k}
={(-1)^{N}\over (k-1)!}\left({\partial\over\partial z}\right)^{k-1}
{1\over (z-\mu_1)\cdots(z-\mu_N)-(-1)^N\lambda}\Bigg|_{z=0},
\ee
\be
\tr\,\Pi\,M^{k}
=-{1\over \Gamma(k-N+2)}\left({\partial\over\partial y}\right)^{k-N+1}
{1\over (y\mu_1-1)\cdots(y\mu_N-1)-y^N\lambda}\Bigg|_{y=0}
\ee
Note that for $k\le N-2$ the last equation gives zero in accordance with
(\ref{102c}).

This leads to a very simple and convenient representation
\be
\tr\,\Pi\,f(M)=(-1)^{N-1}
{1\over 2\pi i}\int\limits_{C_M}dz\,
{f(z)\over (z-\mu_1)\cdots(z-\mu_N)-(-1)^N\lambda}
\label{70e}
\ee
We will need also, in particular, the following formula
\bea
\tr\,\Pi\,e^{-tM}
&=&(-1)^{N-1}
{1\over 2\pi i}\int\limits_{C_M}dz\,
{e^{-tz}\over (z-\mu_1)\cdots(z-\mu_N)-(-1)^N\lambda}
\nonumber\\
&=&{1\over t}{\partial\over\partial\lambda}\tr\,e^{-tM}
=-\sum_{j=1}^N 
{\partial q_j\over\partial\lambda}e^{-tq_j}.
\nonumber\\
\eea
{}From (\ref{101c}) it follows that this function is of order $O(t^{N-1})$ as
$t\to 0$.
{}Using eq. (\ref{83a}) we obtain 
\be
{\partial q_j\over\partial\lambda}
=-\left\{\sum_{k=1}^N
\prod_{i=1;\,i\ne k}^{N}(\mu_i-q_j)\right\}^{-1}.
\ee

In particular case $\lambda=0$ we have 
\be
q_j\Bigg|_{\lambda=0}=\mu_j,
\ee
hence,
\be
{\partial q_j\over\partial\lambda}\Bigg|_{\lambda=0}
=-\prod_{k=1;\,k\ne j}^{N}{1\over (\mu_k-\mu_j)}
\ee
and
\be
\tr\,\Pi\,e^{-tM}\Bigg|_{\lambda=0}
=\sum_{j=1}^N 
e^{-t\mu_j}\prod_{k=1;\,k\ne j}^{N}{1\over (\mu_k-\mu_j)}.
\ee

These formulas will be useful when studying the heat kernel for the operator 
${\cal F}$.

\subsection{Green function of the operator $H-\lambda$}

This shows that the problem is reduced to a Laplace type operator 
${\cal F}$ with an additional matrix structure.
Using the usual heat kernel representation for the Green function ${\cal G}$ 
of the Laplace type operator ${\cal F}$ we have from eq. (\ref{16})
\begin{corollary}
\be
G_{H-\lambda}
=\int\limits_0^\infty dt\,e^{-tm^2}
{\rm tr\,}\Pi\,\exp(-t{\cal F}),
\label{17f}
\ee
or in the kernel form
\be
G_{H-\lambda}=(4\pi)^{-d/2}\Delta^{1/2}
\int\limits_0^\infty dt\,t^{-d/2}\,
\exp\left(-tm^2-{\sigma\over 2t}\right)
\omega_{{\cal F}}(t),
\label{17}
\ee
where
\be
\omega_{{\cal F}}(t)=\tr\,\Pi\,\Omega_{{\cal F}}(t),
\label{17g}
\ee
$\Omega_{{\cal F}}(t)=\Omega_{{\cal F}}(t|x,x')$
is the transfer function of the operator
${\cal F}$ (\ref{800}).
\end{corollary}

Thus we see that there is analogous heat kernel representation for the Green
function
of higher-order operator $H-\lambda$ with a new `transfer function' 
$\omega_{{\cal F}}(t)$.

\subsubsection{Power of a Laplace type operator}

The simplest case is, of course, the case of equal potential terms $Q_j$
and the connections $\nabla^{(i)}$, i.e.
\be
Q_i=Q, \qquad \nabla^{(i)}=\nabla.
\ee
In this case all operators $F_j$ are equal to each other 
\be
F_j=F=-\sq+Q,
\ee
and the operator $H$ is the $N$-th power of the Laplace type operator
$F+m^2$
\be
H=(F+m^2)^N.
\ee
Further, in this simple case we have
\be
{\cal F}=F-P
\ee
and the operator $F$ commutes with the matrix $P$
\be
[F,P]=0.
\ee
Therefore,
\be
\exp(-t{\cal F})=\exp(tP)\exp(-tF).
\ee
and, using eqs. (\ref{16}) and (\ref{200}) we get
\be
\omega_{{\cal F}}(t)=f(t)\Omega_{F}(t),\qquad
f(t)=\tr\,\Pi\exp(tP),
\ee
which
is determined by (\ref{80}) and (\ref{67})
\be
f(t)
={1\over 2\pi i}\int\limits_{C_P} dz {e^{tz}\over z^N-\lambda}
=-{1\over N}(-\lambda)^{-(N-1)/N}
\sum_{j=1}^N e^{i\,\pi(2j-1)/N}e^{tp_j(\lambda)}.
\label{80b}
\ee
On the other hand, expanding the integrand in powers of $\lambda $
and using the formula (\ref{92}) we obtain
$f(t)$ in form of a power series
\bea
f(t)
=t^{N-1}\sum_{k\ge 0}{\lambda^k t^{Nk}\over (Nk+N-1)!}.
\label{67c}
\eea
Thus, the function $\omega_{\cal F}(t)$ is of order $O(t^{N-1})$ as $t\to 0$.

In case $\lambda=0$ we obtain herefrom the usual 
heat kernel representation for the inverse power of the operator
$(F+m^2)$
\be
G_{H}=(F+m^2)^{-N}=
{1\over (N-1)!}\int\limits_0^\infty dt\, t^{N-1}e^{-tm^2}
\exp(-tF).
\ee

\subsubsection{Constant perturbations of the potential terms}

Let us consider now the case when the Laplace type operators $F_j$
differ from each other only by a constant part of the potential terms,
i.e. the connection $\nabla^{(j)}$ is the same, but
the endomorphisms $Q_j$ are different
\be
Q_j=Q+\mu_j,
\ee
so that
\be
F_i=F+\mu_j,
\ee
where
\be
F=-\sq+Q.
\label{11a}
\ee
The operator $H$ has the form
\be
H=(F+m^2+\mu_N)\cdots(F+m^2+\mu_2)(F+m^2+\mu_1).
\ee
The operators $F_i$, even if different, commute.
The corresponding operator ${\cal F}$ has the form
\be
{\cal F}=F+M,
\ee
where $M=\tilde\mu-P$ and
$\tilde\mu=\mu_1\oplus\cdots\oplus\mu_N$ is a diagonal
matrix with the constants $\mu_j$ on the diagonal.
The matrix $M$ commutes with the operator $F$
\be
[F,M]=0.
\ee
Therefore,
\be
\omega_{{\cal F}}(t)=h(t)\Omega_F(t),\qquad
h(t)=\tr\,\Pi\exp(-tM).
\ee
Using the eigenvalues of the matrix $M$, which
are determined by the eq. (\ref{83a}), we obtained this function in the 
form
\bea
h(t)
&=&(-1)^{N-1}
{1\over 2\pi i}\int\limits_{C_M}dz\,
{e^{-tz}\over (z-\mu_1)\cdots(z-\mu_N)-(-1)^N\lambda}
\nonumber\\
&=&-\sum_{j=1}^N 
{\partial q_j\over\partial\lambda}e^{-tq_j}.
\nonumber\\
\eea
As it has been shown in section 3.2, this function is of order $O(t^{N-1})$
as $t\to 0$.
For $\lambda=0$ we obtained 
\be
h(t)\Bigg|_{\lambda=0}
=\sum_{j=1}^N
\left[\prod_{k=1;\,k\ne j}^N{1\over (\mu_k-\mu_j)}\right]
e^{-t\mu_j}.
\ee

\subsubsection{General case}

As we have seen, in particular cases studied before the
function $\omega_{{\cal F}}(t)$ defined by
(\ref{17g}) is of order $O(t^{N-1})$
as $t\to 0$. It is almost clear that this is valid also in general, since the
main contribution to
the asymptotics give the constant background fields.
This can be formulated in form of a lemma.

\begin{lemma}
Let ${\cal F}$ be the Laplace type operator defined by (\ref{800}) and
$\Pi$ be the matrix defined by (\ref{17a}).
Then for its HMDS-coefficients $a_k({\cal F})$ there holds
\be
\tr\,\Pi\,a_{k}({\cal F})=0 \qquad {\rm for}\ k=0,1,\dots, N-2.
\ee
In other words, for the asymptotic expansion as $t\to 0$ of 
its transfer function $\omega_{{\cal F}}(t)$ there holds
\be
\omega_{{\cal F}}(t)\Bigg|_{t\to 0}
\sim{(-t)^{N-1}\over (N-1)!}\sum_{k\ge 0}{(-t)^k\over k!}c_k({\cal F}),
\ee
where
\be
c_k({\cal F})={(N-1)!k!\over(N+k-1)!}\tr\,\Pi\,a_{k+N-1}({\cal F}).
\ee

\end{lemma}  
\paragraph{Proof.}\
The coefficients $a_k$ can be calculated in form of a covariant Taylor series
near the diagonal (for details and notation see \cite{avr91b})
\be
a_k=\sum_{n\ge 0}|n><n|a_k>,
\ee
where $|n>$ is the covariant Taylor basis and $<n|a_k>$ are the diagonal values
of
the symmetrized covariant derivatives. 
$<n|a_k>$ have the following general form \cite{avr91b}
\be
<n|a_k>=\sum_{n_1,\dots,n_{k-1}\ge 0}
c(n_1,\dots, n_{k-1})
<n\vert M\vert n_{k-1}><n_{k-1}\vert M\vert n_{k-2}>\cdots<n_1\vert M\vert 0>,
\label{147f}
\ee
where $c(n_1,\dots, n_{k-1})$ are some constants
and $<m|M|n>$ are some ${\rm End}\,(V)$-valued
tensors that vanish for $n>m+2$ and for $n=m+1$.
Important for us is that, since the matrix $P$ is
constant, it appears only in
$<n|M|n>$
\be
<n|M|n>=P+B,
\label{153g}
\ee
where $B$ is a diagonal matrix.
{\it All } others tensors $<n|M|m>$ do not depend on the matrix
$P$ and are diagonal---they are actually polynomials in the curvatures and
their covariant
derivatives \cite{avr91b}.
So, all the terms in the sum (\ref{147f}) have the form 
\be
B_{1}P^{n_1}\dots B_{j}P^{n_j},\qquad j=1,2,\dots,k,
\ee
where $B_i$ are some diagonal matrices and 
the total power of the matrix $P$ is restricted by $n_1+\cdots +n_j\le k$.
Therefore, from (\ref{91f}) we find that $\tr\Pi<n|a_k>=\tr\,\Pi\,a_k=0$ for
$k\le N-2$,
which proves the lemma.

Actually, using (\ref{153g}) and (\ref{70}) one can prove more, namely, 
\be
c_{0}({\cal F})=\tr\,\Pi\,a_{N-1}({\cal F})
=\tr\Pi\,P^{N-1}=1.
\ee

\subsubsection{Hadamard expansion of the Green function
of the operator $H-\lambda$}

The behavior of the Green function of Laplace type operators
near diagonal is studied in the previous section.
So, by using the eqs. of sect. 2 we obtain from (\ref{16})
the Hadamard series
for the Green function of the higher order operator $H-\lambda$
\be
G_{H-\lambda}=G^{\rm sing}_{H-\lambda}
+G^{\rm non-anal}_{H-\lambda}
+G^{\rm reg}_{H-\lambda},
\label{107}
\ee
where the singular part reads
\be
G^{\rm sing}_{H-\lambda}=
(4\pi)^{-{d\over 2}}{\Delta^{1/2}\over (N-1)!}
\sum_{k=0}^{\left[{(d+1)\over 2}\right]-N-1}(-1)^{N+k}
{\Gamma\left({d\over 2}-k-N\right)\over k!}
\left({2\over\sigma}\right)^{{d\over 2}-k-N}
c_{k}({\cal F}+m^2).
\label{108}
\ee
Further, we get in odd dimensions: 
\be
G^{\rm non-anal}_{H-\lambda}\sim
(-1)^{d-1\over 2}(4\pi)^{-{d\over 2}}{\Delta^{1\over 2}\over (N-1)!}
\sum_{k=0}^{\infty}
{\pi \left({\sigma/ 2}\right)^{k+1/2}c_{k-N+{(d+1)/2}}({\cal F}+m^2)
\over \Gamma\left(k-N+1+{d+1\over 2}\right)\Gamma\left(k+{3\over 2}\right)}
\ee
\be
G^{\rm reg}_{H-\lambda}\sim
(-1)^{{(d+1)\over 2}}(4\pi)^{-{d\over 2}}{\Delta^{1/2}\over (N-1)!}
\sum_{k=0}^{\infty}{\pi \left({\sigma/2}\right)^{k}\over k!\Gamma(k-N+1+d/2)}
\varphi_{{\cal F}+m^2}(k-N+d/2),
\ee
and in even dimensions:
\be
G^{\rm non-anal}_{H-\lambda}\sim
(-1)^{d/2-1}(4\pi)^{-d/2}{\Delta^{1/2}\over (N-1)!}
\log\left(\mu^2\sigma\over 2\right)
\sum_{k=0}^{\infty}
{\left({\sigma/2}\right)^{k}
c_{k-N+d/2}({\cal F}+m^2)
\over k!\Gamma(k-N+1+{d\over 2})}
\label{597e}
\ee
\bea
G^{\rm reg}_{H-\lambda}&\sim &
(-1)^{d/2-1}(4\pi)^{-d/2}{\Delta^{1/2}\over (N-1)!}\sum_{k=0}^{\infty}
{1\over k!\Gamma(k-N+1+d/2)}\left({\sigma\over 2}\right)^{k}
\nonumber\\[11pt]
&\times&
\left\{\varphi'_{{\cal F}+m^2}(k-N+d/2)
-\left[\log \mu^2+\Psi(k+1)
+\Psi(k+d/2)\right]c_{k-N+{d\over 2}}({\cal F}+m^2)\right\}
\nonumber\\
&&
\label{112}
\eea
where
\be
\varphi_{{\cal F}+m^2}(q)\equiv {(N-1)!\Gamma(q+1)\over\Gamma(q+N)}
\tr\,\Pi\,b_{{\cal F}+m^2}(q+N-1).
\ee

Thus, we find that
\begin{itemize}

\item[i)]
the structure of the diagonal singularities of the 
Green function $G_{H-\lambda}$
is determined by the coeffcieints $c_k({\cal F}+m^2)$;

\item[ii)]
in odd dimenions there is no logarithmic singularity;

\item[iii)]
{}for $N>d/2$ 
there are {\it no} singularities at all and the Green function is regular
on the diagonal, i.e. there exists finite coincidence limit $G^{\rm
diag}_{H-\lambda}\equiv 
G_{H-\lambda}(x,x)=G^{\rm reg}_{H-\lambda}(x,x)$;

\item[iv)]
{}for $N=\left[(d+1)/2\right]$ there are no singularities for odd dimension 
$d=2N-1$ and there 
is only logarithmic singularity in even dimension $d=2N$.

\item[v)]
The condition for the validity of the Huygens principle for
the operator $H-\lambda $
in even dimensions reads
\be
\sum_{k=0}^{\infty}
{1\over k!\Gamma(k-N+1+d/2)}\left({\sigma\over 2}\right)^{k}
c_{k-N+d/2}({\cal F}+m^2)=0.
\ee
Similarly to (\ref{597g}) this also produces an infinite set of local
conditions.

\end{itemize}

\subsection{Heat kernel of the operator $H$}

The heat semigroup for the operator $H$ 
can be obtained by using the standard formula
\bea
\exp(-tH)&=&-{1\over 2\pi i}\int\limits_{C_H} d\lambda
e^{-t\lambda}G_{H-\lambda}
\nonumber\\
&=&{t^{-k}\over 2\pi i}\int\limits_{c-i\infty}^{c+i\infty}
d\lambda
e^{-t\lambda}\left({\partial\over\partial\lambda}\right)^k
G_{H-\lambda},
\eea
where the contour of integration $C_H$ goes counter-clockwise around
the spectrum of the operator $H$, $k$ is some sufficiently large
integer and
$c$ is a negative constant.
Since the operator $H$ is a positive self-adjoint operator
its spectrum lies on the positive real half-axis, i.e. the contour $C_H$ should
go 
from $i\varepsilon+\infty$ to $-i\varepsilon+\infty$ enclosing the origin.
Using the Theorem 1 one can express the heat semigroup for the higher order
operator $H$ in terms of the heat semigroup of the Laplace type
operator ${\cal F}$.
We have first
\be
\exp(-tH)={t^{-k}\over 2\pi i}\int\limits_{c-i\infty}^{c+i\infty} d\lambda\,
e^{-t\lambda}\left({\partial\over\partial\lambda}\right)^k
\tr \Pi\, G_{{\cal F}+m^2},
\label{56}
\ee
Now, using the heat kernel representation of the Green operator
$G_{{\cal F}+m^2}$ we obtain
\begin{corollary}
The heat semigroup for the operator $H$ is determined by the heat semigroup
for the operator ${\cal F}$
\be
\exp(-tH)={t^{-k}\over 2\pi i}\int\limits_{c-i\infty}^{c+i\infty} d\lambda\,
e^{-t\lambda}\int\limits_0^\infty ds\, e^{-sm^2}
\left({\partial\over\partial\lambda}\right)^k\tr\,\Pi\,\exp(-s{\cal F})
\ee
and, consequently, the heat kernel for the operator $H$ is given by
\bea
U_{H}(t)&=&(4\pi)^{-d/2}\Delta^{1/2}
{t^{-k}\over 2\pi i}\int\limits_{c-i\infty}^{c+i\infty} d\lambda\,
e^{-t\lambda}\int\limits_0^\infty ds\, s^{-d/2}\,
\exp\left(-sm^2-{\sigma\over 2s}\right)
\left({\partial\over\partial\lambda}\right)^k\omega_{{\cal F}}(s).
\nonumber\\
\label{162g}
\eea
\end{corollary}

We studied the expansion of the Green function of a Laplace type operator in
previous 
section in detail. The important fact is, that the HMDS-coefficients 
$a_k({\cal F}+m^2)$ for the Laplace type operator $({\cal F}+m^2)$ 
are polynomial in $\lambda$, they are, of course, also polynomial in $m^2$.
This means that the singular part of the Green function (including the
logarithmic 
singularity), which are expressed in terms of $a_k({\cal F}+m^2)$, are also
polynomial in
$\lambda$. On the other hand, it is well known that
\be
-{1\over 2\pi i}\int\limits_{C_H} d\lambda\,
e^{-\lambda}(-\lambda)^{z}={1\over\Gamma(-z)}.
\label{92}
\ee
{}For integer $z=k=0,1,2,\dots$ we have, therefore,
\be
-{1\over 2\pi i}\int\limits_{C_H} d\lambda\,
e^{-\lambda}(-\lambda)^{k}=0.
\ee

Thus, the singular part, which is polynomial in $\lambda$, 
does not contribute to the integral (\ref{56}). 
Therefore, one can substitute in this equation instead of the whole Green
function only 
the regular part of it 
$G^{\rm reg}$, go to the diagonal and take the trace 
\be
\Tr_{L^2}\exp(-tH)={t^{-k}\over 2\pi i}\int\limits_{c-i\infty}^{c+i\infty}
 d\lambda\,e^{-t\lambda}
\left({\partial\over\partial\lambda}\right)^k
\Tr_{L^2}\tr\Pi\,G^{\rm reg}_{{\cal F}+m^2},
\label{56e}
\ee
Finally, using the eqs. (\ref{25}) and (\ref{33}) we have:
in odd dimensions
\be
\Tr_{L^2}\exp(-tH)=
{(-1)^{(d+1)/2}(4\pi)^{-d/2}\pi\over (N-1)!\Gamma(d/2-N+1)}
{t^{-k}\over 2\pi i}\int\limits_{c-i\infty}^{c+i\infty} d\lambda\,
e^{-t\lambda}
\left({\partial\over\partial\lambda}\right)^k
\Phi_{{\cal F}+m^2}(d/2-N),
\label{56a}
\ee
and in even dimensions
\be
\Tr_{L^2}\exp(-tH)=
{(-1)^{d/2-1}(4\pi)^{-d/2}\over (N-1)!\Gamma(d/2-N+1)}
{t^{-k}\over 2\pi i}\int\limits_{c-i\infty}^{c+i\infty} d\lambda\,
e^{-t\lambda}
\left({\partial\over\partial\lambda}\right)^k
\Phi'_{{\cal F}+m^2}(d/2-N),
\label{56b}
\ee
where $\Phi_{{\cal F}+m^2}(q)=\Tr_{L^2}\varphi_{{\cal F}+m^2}(q)$
and $\Phi'_{{\cal F}+m^2}(q)=(d/dq)\Phi_{{\cal F}+m^2}(q)$.

These formulas may be used, for example, to compute the
asymptotic expansion of the functional trace
of the heat kernel for the higher order operator $H$, which is known to have
the following general form \cite{gilkey84}
\be
\Tr_{L^2}\exp(-tH)\sim 
(4\pi)^{-d/2}{\Gamma\left[{d/(2N)}\right]\over
N\Gamma\left({d/2}\right)}\,
t^{-{d/(2N)}}\sum_{k\ge 0}t^{k/N}E_k(H).
\ee
We normalize the coefficients $E_k(H)$ so that $E_0(H)=
{\rm Tr}_{L^2}\,1$. For $N=1$ the coefficients $E_k$ are determined by the
HMDS-coefficients
$E_k=[(-1)^k/k!]A_k$.
However, one should note that to compute the coefficients $E_k(H)$ one
needs to know the asymptotics of the function
$\omega_{\cal F}(t)$
(or the functions $\Phi_{{\cal F}+m^2}(d/2-1)$ and
$\Phi'_{{\cal F}+m^2}(d/2-1)$)
as $\lambda\to-\infty$, which is unknown, in general,
in terms of the known asymptotic expansion as $t\to 0$, i.e. in terms
of the HMDS-coefficients for a Laplace type operator.
The known asymptotics of these functions
as $s\to 0$ and $m\to\infty$ do not contribute at all
in the eqs. (\ref{162g}), (\ref{56a}) and (\ref{56b}).

\section*{ACKNOWLEDGMENTS}

I am greatly indebted to Peter Gilkey and Rainer Schimming for stimulating
and fruitful discussions.
This work was supported by the Deutsche Forschungsgemeinschaft.


\end{document}